# Accurate Multi-physics Numerical Analysis of Particle Preconcentration Based on Ion Concentration Polarization

Zirui Li,[1*] Wei Liu,[1] Yudan Zhu,[2] Xiaohua Lu,[2] Yuantong Gu,[3] Jongyoon Han[4,5,6]

[1] College of Mechanical and Electrical Engineering, Wenzhou University, Wenzhou, China.
[2] College of Chemical Engineering, State Key Laboratory of Materials-oriented Chemical Engineering, Nanjing Tech University, Nanjing, China.
[3] School of Chemistry, Physics and Mechanical Engineering, Queensland University of Technology, Australia
[4] Department of Biological Engineering, Massachusetts Institute of Technology, Cambridge, Massachusetts, USA.
[5] Department of Electrical Engineering and Computer Science, Massachusetts Institute of Technology, Cambridge, Massachusetts, USA.
[6] Singapore-MIT Alliance for Research and Technology, Singapore.

**ABSTRACT**

This paper studies mechanism of preconcentration of charged particles in a straight micro-channel embedded with permselective membranes, by numerically solving coupled transport equations of ions, charged particles and solvent fluid without any simplifying assumptions. It is demonstrated that trapping and preconcentration of charged particles are determined by the interplay between drag force from the electroosmotic fluid flow and the electrophoretic force applied trough the electric field. Several insightful characteristics are revealed, including the diverse dynamics of co-ions and counter ions, replacement of co-ions by focused particles, lowered ion concentrations in particle enriched zone, and enhanced electroosmotic pumping effect *etc*. Conditions for particles that may be concentrated are identified in terms of charges, sizes and electrophoretic mobilities of particles and co-ions. Dependences of enrichment factor on cross-membrane voltage, initial particle concentration and buffer ion concentrations are analyzed and the underlying reasons are elaborated. Finally, *post priori* a condition for validity of decoupled simulation model is given based on charges carried by focused charge particles and that by buffer co-ions. These results provide important guidance in the design and optimization of nanofluidic preconcentration and other related devices.

## I. INTRODUCTION

Electrokinetic manipulation for electrolytes and particles in micro- and nano-scale fluidic systems experienced significant progresses in the last decade[1]. While electroosmotic flow (EOF) has been well-characterized in microfluidics[2], transport behaviors of ion and fluids in nanoscale channels or pores are much less understood[3]. Overlapping electric double layers (EDLs) generates a lot of new physics and invoked many novel applications. Basically, inside nanoscale spaces, the amount of counter ions is significantly greater than that of co-ions[4]. When a nanochannel is used to bridge two microchannels and an electric field is applied along the channel axis (as shown in Fig. 1a), the number of counter ions transported is significantly greater than that of co-ions, rendering ion current permselective. This selective charge transport results in ion depletion and enrichment zones in either side of the nanochannel, *a.k.a.* ion concentration polarization (ICP). Such micro-nano-fluidic systems facilitate many unique and novel functionalities, such as desalination[5-7], particle preconcentration[8-10] and biomolecular separation *etc.* [11,12]. Because materials containing nanoscale pores with charged surfaces have the same ion exclusion effects, permselective membranes (*e.g.* Nafion) are widely used to substitute nanochannels, eliminating the need for more involved nanoscale fabrications.

---

[*] lizirui@wzu.edu.cn

Physics underlying nanofluidic ICP is complicated[13]. It involves coupled nonlinear fluid flow, ionic transport and dynamic evolution of electric potential. In addition to ICP, selective transport of ions across nanochannels induces a strong electric field and an extended space charge layer at the depletion side, which produces a fast, vortical, sometimes even chaotic fluid flow under external electric field[14]. Study of such complicated systems has been largely based on experimental observation of tracing particles. Unfortunately however, direct experimental monitoring of some key parameters (*e.g.* ion concentrations) in such micro-nano-fluidic systems is still impractical, prohibiting accurate understanding of the system. Precise description of mechanisms has to rely on numerical simulation. In this aspect, there have been some simulation studies on ICP propagation[15-17], vortical flow[18-21], and desalination of solution of binary ions[22]. However, very few simulation studies have been reported for preconcentration processes. Recently, Shen *et al.* simulated EOF driven protein preconcentration process in a microchannel integrated with Nafion strip[23]. They used constant zeta potential to describe surface charge effect and did not consider the effect of (accumulated) charged proteins on the distribution of electric field. These assumptions will become invalid when high variation of electric field is involved and/or high level of charged protein concentration is achieved, as often encountered in preconcentration experiments for low abundance proteins. Wang *et al.* conducted a more accurate simulation for pressure-driven particle preconcentration and established dependence of the focusing rate upon diffusivity and electrophoretic mobility of particles[24]. Yet, their simple model does not involve coupled fluid flow and is limited to extremely slow flow conditions.

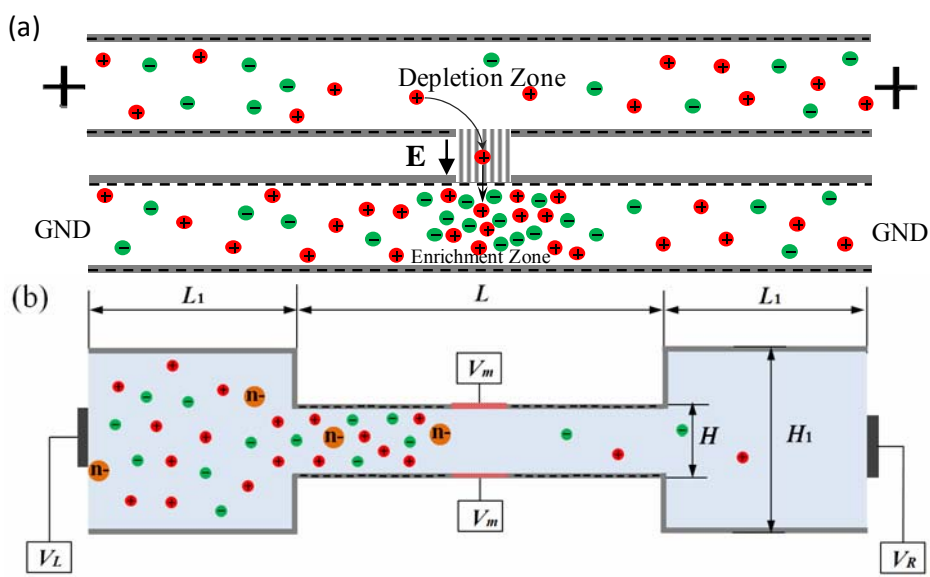

FIG. 1. (a) Schematic sketch of ICP systems. (b) Physical setup of particle preconcentration system.

The purpose of this paper is to study the mechanism of charged particle preconcentration through accurate multi-physics modeling without any simplifying assumptions. Mechanisms underlying particle enrichment will be elaborated in terms of interplay between the viscous drag force and the electrophoretic force applied on the charged particles. Conditions for particle enrichment will be identified. By analyzing eventual steady states of the system, replacement of co-ions by the focused particles and its consequences will be elaborated. Finally, a dynamic picture for transport of ions, particles and fluids are given, to offer a new way of explaining experimental data and guiding the design of relevant systems.

## II. METHODS

### A. System setup

Fig. 1b shows a schematic of two-dimensional (2D) model for particle preconcentration simulation. The key component of the system is a microchannel of length $L=120\mu m$ and width $H=4\mu m$ embedded with permselective membranes of length $L_m=2\mu m$ in the middle of both top and bottom walls. Connected to the microchannel are two wider channel segments of width and width $L_1=H_1=60\mu m$ (simulating part of the reservoirs). The microchannel walls are charged with density $\sigma_-=-5mC/m^2$, while walls of wider channels are uncharged. The left boundary is connected to mixed solution of potassium chloride (KCl) and negatively charged particles (denoted by $P^{n-}$, with $n$ representing the valence number). An electric field is applied such that voltage at the left boundary is $V_L=20V_T$ ($V_T=25.8mV$ is the thermal voltage) and that at the right boundary is $V_R=0$, inducing an EOF from left to right. Pressures at both the left and the right boundaries are set to zero such that transport of ions and fluids is resulted from electrokinetic phenomena only. The permselective membranes are assumed to permit passage of $K^+$ only. Voltages at the membranes surface are independently set to $V_m$. For convenience, a term *cross membrane voltage* $V_{cm}=(V_L+V_R)/2-V_m$ is defined as the difference between voltage at the membrane location determined by external electric field without ICP and the actual voltage applied on the membrane, representing the voltage drop across the nanochannel/membrane junction in actual experiments. Lower $V_m$ corresponds to larger $V_{cm}$ and generates stronger ion depletion effect.

### B. Governing equations

Governing equations for the incompressible fluid flow, ion and particle transport, and electric potential are described by the Navier-Stokes, Nernst-Planck, and Poisson equations, respectively[25,26]:

$$\rho\left(\partial \mathbf{U}/\partial t+(\mathbf{U}\cdot\nabla)\mathbf{U}\right)=-\nabla P+\eta\nabla\cdot\nabla\mathbf{U}-\rho_e\nabla\Phi, \qquad (1)$$

$$\nabla\cdot\mathbf{U}=0 \qquad (2)$$

$$\frac{\partial C_i}{\partial t}=-\nabla\cdot\mathbf{J}_i, \qquad (3)$$

$$\mathbf{J}_i = -\left(D_i \nabla C_i + Z_i (D_i F / RT) C_i \nabla \Phi\right) + \mathbf{U} C_i. \tag{4}$$

$$-\nabla \cdot (\varepsilon \nabla \Phi) = \rho_e. \tag{5}$$

Here, $\mathbf{U}$ is the velocity, $P$ is the pressure, $C_i$ and $\mathbf{J}_i$ are the concentration and flux density of species $i$, respectively. For convenience, we use $i = 1$ for K$^+$, $i = 2$ for Cl$^-$, and $i = 3$ for P$^{n-}$. $Z_i$ and $D_i$ are the valence and diffusion coefficient of species $i$ respectively. $\Phi$ is the electric potential, $\rho_e = e \sum_{i=1}^{3} Z_i C_i$ is the free space charge density, with $e$ being the elementary charge. Parameters $\rho$, $\eta$ and $\varepsilon$ are the mass density, dynamic viscosity and permittivity of the solution respectively. $T$ is absolute temperature. Constants $F$ and $R$ are Faraday's number and gas constant respectively.

### C. Boundary conditions

At the membrane's surfaces it is assumed that[25,26]: (i) Fluxes of anions and particles across the membrane are zero; (ii) The concentration of cations at the membrane surface is $2 \cdot C_0$; (iii) The electric potential at the membrane surface is $V_m$; (iv) The membrane surface is impermeable and no-slip to fluid, indicating zero velocity and zero-gradient condition for the pressure. The corresponding equations are:

$$\mathbf{J}_2 \cdot \mathbf{n} = 0, \ \mathbf{J}_3 \cdot \mathbf{n} = 0, \ C_1 = 2 \cdot C_0, \ \Phi = V_m, \ \mathbf{U} = \mathbf{0}, \ \nabla P \cdot \mathbf{n} = 0. \tag{6}$$

At microchannel walls the boundary conditions are: (i) constant surface charge density; (ii) no-slip condition for fluid velocity and zero-gradient conditions for pressure; (iii) impermeability to ions and particles:

$$\sigma_{-} = -5 \text{mC/m}^2, \ U = 0, \ \nabla P \cdot \mathbf{n} = 0, \ \mathbf{J}_i \cdot \mathbf{n} = 0, \ i = 1,2,3. \tag{7}$$

At the inlet boundary: (i) Both concentration of electrolytes and particles are the same as those in the inlet reservoir. (ii) The electric potential is $V_L$; (iii) The pressure is zero:

$$\Phi = V_L, \ \nabla U \cdot \mathbf{n} = 0, \ P = 0, \ C_i = C_{i,0}, \ i = 1,2,3. \tag{8}$$

At outlet boundary: (i) *free* boundary conditions are applied for fluid flow; (ii) the electric potential is set to $V_R$.

$$\Phi = V_R, \ \nabla U \cdot \mathbf{n} = 0, \ P = 0, \ \nabla C_i \cdot \mathbf{n} = 0, \ i = 1,2,3. \tag{9}$$

At reservoirs walls the boundary conditions are: (i) no-slip condition for fluid velocity and zero-gradient conditions for pressure; (ii) Zero Charge.

$$U = 0, \ \nabla P \cdot \mathbf{n} = 0, \ \nabla \Phi \cdot \mathbf{n} = 0. \tag{10}$$

### D. Numerical methods

In this computation, the governing equations are solved with the specified boundary conditions using COMSOL v5.2a (a full model is provided as supplemental information). Based on symmetries in geometry and physics setup, only the lower half of the channel is modeled with symmetric boundary conditions at the upper boundary of the calculation model (the central line of the channel). The computational domain is meshed using quadrilateral elements. Finer grids are adopted near the charged wall, membrane surface, inlet and outlet boundaries of the channel. Concentration and potential fields are implemented in physics interfaces of Transport of Diluted Species and Electrostatics interface. The Poisson-Nernst-Planck (PNP) equations are solved using quadratic Lagrange interpolation functions for space discretization. Navier–Stokes (NS) and continuity equations are implemented in Creeping Flow interface. Quadratic Lagrange shape functions are used for NS equations whereas linear functions are used for the continuity equation.

For time dependent analysis, we adopt two different configurations at different stages of simulation. At the beginning stage ($t \leq 0.5\text{s}$), PNP and NS equations are solved segregatedly [27]. At each time step, the velocity and pressure are solved first (with electric force calculated from the charges at the previous step), followed by concentrations and potentials (with the velocity and pressure obtained at this step). Default setting of multifrontal massively parallel sparse direct solver (MUMPS) is adopted with an implicit generalized $\alpha$ method ($\alpha = 0.99$). In this stage, ion-depletion zone emerges and vortices are generated. After that, fully coupled PNP and NS equations are solved using backward differentiation formula (BDF) method with free time steps, until the end of simulation time.

For steady state analysis, fully coupled PNP and NS equations are solved using default solver (automatic Newton nonlinear method). Due to the strong nonlinear characteristics of the system and the inconsistency in the initial conditions for concentration and electric potential near the channel or membrane surfaces, it is necessary to ramp up the membrane voltages slowly to the desired value. This is implemented through auxiliary sweep of parameters within the solver.

## III. RESULTS AND DISCUSSIONS

### A. Developments of particle concentration and concentration speed

We start with a time-dependent simulation for enrichment of divalent particles P$^{2-}$ ($Z_3 = -2$, $n = 2$) of diameter 4 times larger than Cl$^-$ ($D_3 = D_2/4$, because diffusion coefficient of a spherical particle is inversely proportional to its diameter) under $V_{cm} = 30V_T$. The initial concentrations are the same as those in the reservoir ($C_{1,0} = 1\text{mM}$, $C_{3,0} = 10^{-7}\text{mM}$, and $C_{2,0} = C_{1,0} - 2C_{3,0}$). This setup guarantees that the system runs in a diluted solution regime, which does not involve correction of fluid properties, such as density and viscosity, even when the particles are focused (actually, the focused particles will replace co-ions and reduces the concentration of buffer ions, making the solution diluter, c.f. Section E). Fig. 2a shows the evolution of particle concentration inside the microchannel (colors not to scale, concentration values can be found in the corresponding curves in Fig. 2b). It could be found that shortly after application of $V_{cm}$ (e.g. $t = 6\text{ms}$), particles begin to accumulate in the upstream microchannel and the enrichment factor $r = C_3/C_{3,0}$ grows. In the mean time, particle concentration near the membrane location is lowered and this low concentration region expands to the downstream with the fluid flow. At $t = 2\text{s}$, the enrichment factor in the upstream channel reaches $\sim 10^2$ and the concentration in the downstream microchannel is the lowest. The enrichment factor exceeds $10^3$, $10^4$ and $10^5$ at times $14\text{s}, 200\text{s}$, and $1800\text{s}$ respectively, accompanied by increasing of particle concentration in the downstream. This is because the number of particles overcoming the electric energy barrier at the front of ion depletion zone increases when more particles are accumulated before the barrier. After infinitely long time ($t = \infty$), the system should approach steady state, which could be obtained numerically through stationary analysis (see the dashed curve in Fig. 2b). At this state, the number of particles entering the channel through the left boundary is equal to that leaving the channel through the right boundary. The amount of particles inside the channel reaches a constant limiting value, with a million-fold enrichment of particle concentration ($r = 3.2 \times 10^6$).

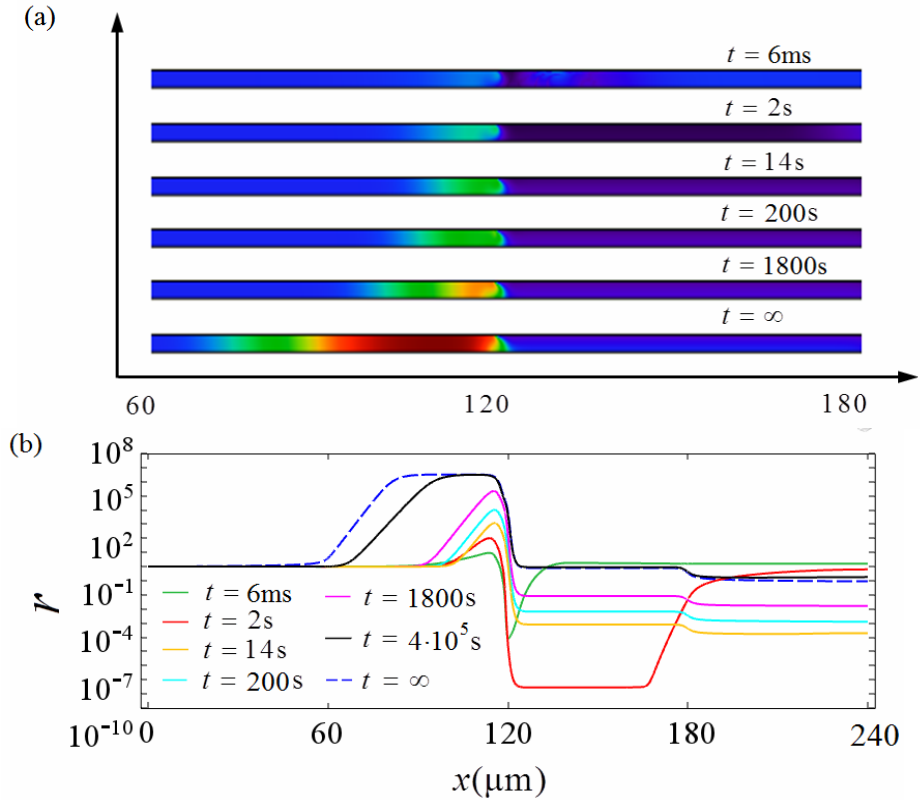

FIG 2. Concentration of $P^{2-}$ at different times. (a) 2D distributions (colors not to scale); (b) Profiles of enrichment factor $r$ along the center line of the channel.

It has been repeatedly observed in experiments that concentration speed in such micro-nano-fluidic system exhibits a complicated, nonlinear characteristics: at the beginning, the concentration speed increases; in the second stage, it is almost a constant; and finally, it decreases slowly. To experimentalists, this *concentration speed* is largely approximate, estimated based on the peak concentrations measured through the fluorescence intensity as the metric and the width of the peak. A more rigorous definition of the concentration speed could be defined though the total amount of particles trapped in the upstream microchannel. This corresponds to the time derivative of the integration of particle concentration over the upstream microchannel in Fig 2a, or approximately the difference in fluxes at the inlet and outlet boundaries of the microchannel, *i. e.* $j_{x,3}|_{\text{inlet}} - j_{x,3}|_{\text{outlet}}$ (this include the amount of particles in the downstream microchannel, which is negligibly small). Fig. 3 shows the evolution of integrated particle fluxes over the inlet and outlet boundaries of the microchannel, as well as the difference between them. From the solid black curve, we may find that the concentration speed grows by approximately 50% in about 20 seconds (stage I), then it is followed by an almost constant concentration speed for a long time (stage II ). At times beyond $10^4$s the concentration speed decreases gradually to almost zero at time $t = 4 \cdot 10^5$s (stage III), after that, the system approaches the steady state (stage IV). In Stage I, the depletion zone, the electric energy barrier, and the vortex driven electrokinetic flow are developing, bring particles into the microchannel at an increasing speed. After the electric field and the electrokinetic becomes stable, the system enters into Stage II with constant feeding rate of the particles. In these two stages, the flux out of the microchannel is negligible compare to that enters into the microchannel, with almost all the particles focused. Therefore, an increasing concentration speed in stage I and a constant concentration speed in stage II are observed. However, in stage III, the concentration of particles reaches maximum limiting value and the focused peak expands into the upstream. The flux at the inlet is reduced while the flux at the outlet is increased by strong diffusion of particles into both ends of the channel. As a results, the concentration speed reduces. Ultimately, the number of the particles that enters the microchannel at a unit time become equal to that leaves the microchannel at the outlet, and the system reaches a steady state, with zero concentration speed (stage IV). Although this limiting steady state often needs quite long time to reach, it defines unique behaviors and ultimate functionalities of the system. Thus we use such steady states for further analyses.

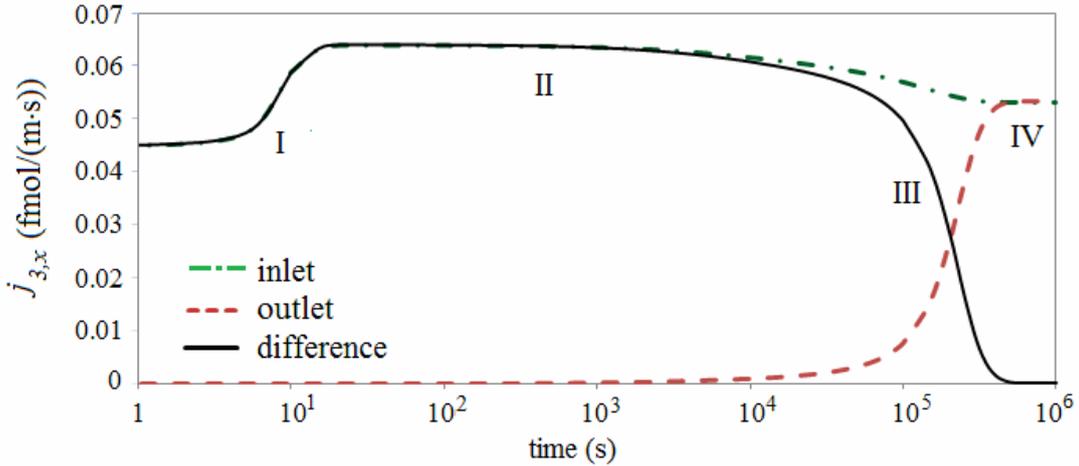

FIG. 3. Evolution of concentration speed (solid black curve) expressed as the different between the fluxes at the inlet (dotted dashed, green) and outlet (dashed, red) boundaries.

## B. Effects of cross membrane voltage

As ion depletion and particle enrichment are induced by cross membrane voltage $V_{cm}$, it is expected that higher $V_{cm}$ will induce stronger enrichment effect. This tendency is demonstrated in the profiles of particle concentration shown in Fig. 4. Here, the parameters of the particles are the same as those in Fig. 2 ($Z_3 = -2$, $D_3 = D_2/4$), while $V_{cm}$ is varied in a wide range. It could be found that the enrichment factor increases sharply with $V_{cm}$ when $V_{cm}$ is smaller than $26V_T$, but it becomes saturated (at an value of $r = 3.2 \times 10^6$) beyond that value. Further increasing of $V_{cm}$ does not increase the height of the peak, but broadens it into the upstream. The inset of Fig. 4 shows the relationship between the maximum enrichment factor and $V_{cm}$. It could be found that there are two distinct phases in the curve. For $V_{cm} < 26V_T$, the enrichment factor grows exponentially with $V_{cm}$. In this phase, increasing of $V_{cm}$ strengthens the electric field approximately linearly, which results in an exponentially increasing concentration of particles at the steady state (See Eq. 4). However, for $V_{cm} > 26V_T$, the concentration of particles becomes so high that they replace almost all the co-ions in the solution. In this scenario, the particle concentration is limited by the concentration of counter ions (decreasing with $V_{cm}$, c.f. Section E), and therefore the enrichment factor becomes saturated.

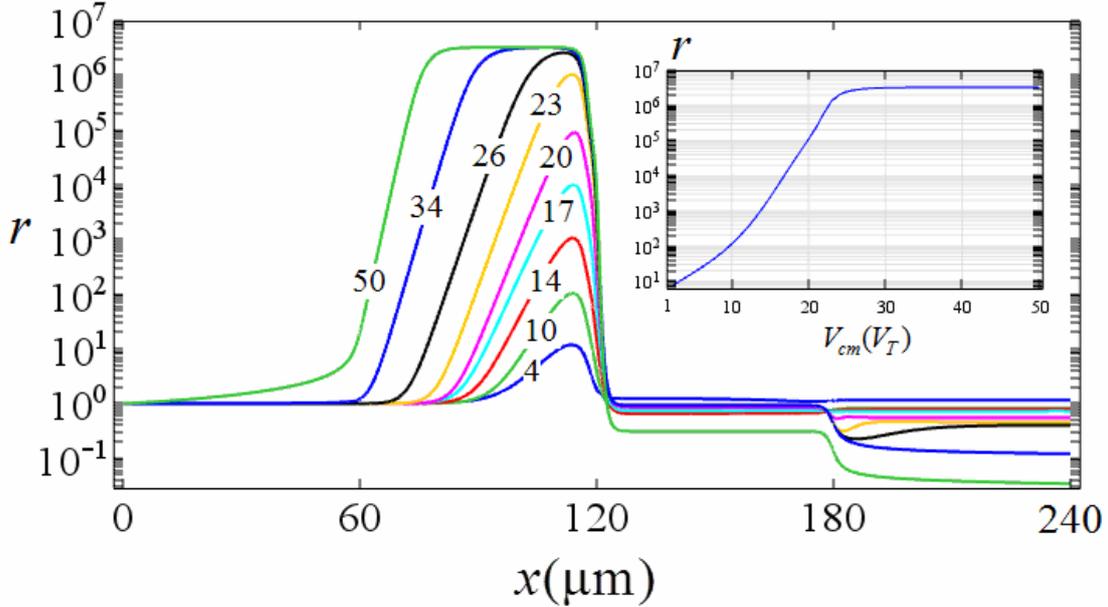

FIG. 4. Steady state profiles of particle concentration along the center line of the microchannel under varied $V_{cm}$ (labels on the curves are $V_{cm}$ in terms of $V_T$). Inset shows dependence of the maximum enrichment factor on $V_{cm}$.

### C. Effects of initial particle and buffer ion concentrations

For a given channel, apart from $V_{cm}$, two most important factors that may affect the enrichment of particles are the initial particle concentration in the inlet reservoir ($C_{3,0}$) and the concentration of buffer ions ($C_0$). To investigate these effects, we calculate the limiting concentration of charged particles with varied $C_{3,0}$ and $C_0$ respectively. Fig. 5a shows the result under $V_{cm} = 30V_T$, from which we learn that the enrichment factor is generally inversely proportional to $C_{3,0}$, which implies a constant limiting particle concentration inside the microchannel. In fact, the maximum particle concentration (shown in the inset) is very slightly dependent on particle concentration in the inlet reservoir. Specifically, when $C_{3,0}$ increases from $10^{-7}$ mM to $10^{-2}$ mM (by an order of $10^5$), the limiting focused concentration grows only from 0.32mM to 0.35mM (by ~10%). Such a variation is practically negligible in most engineering applications.

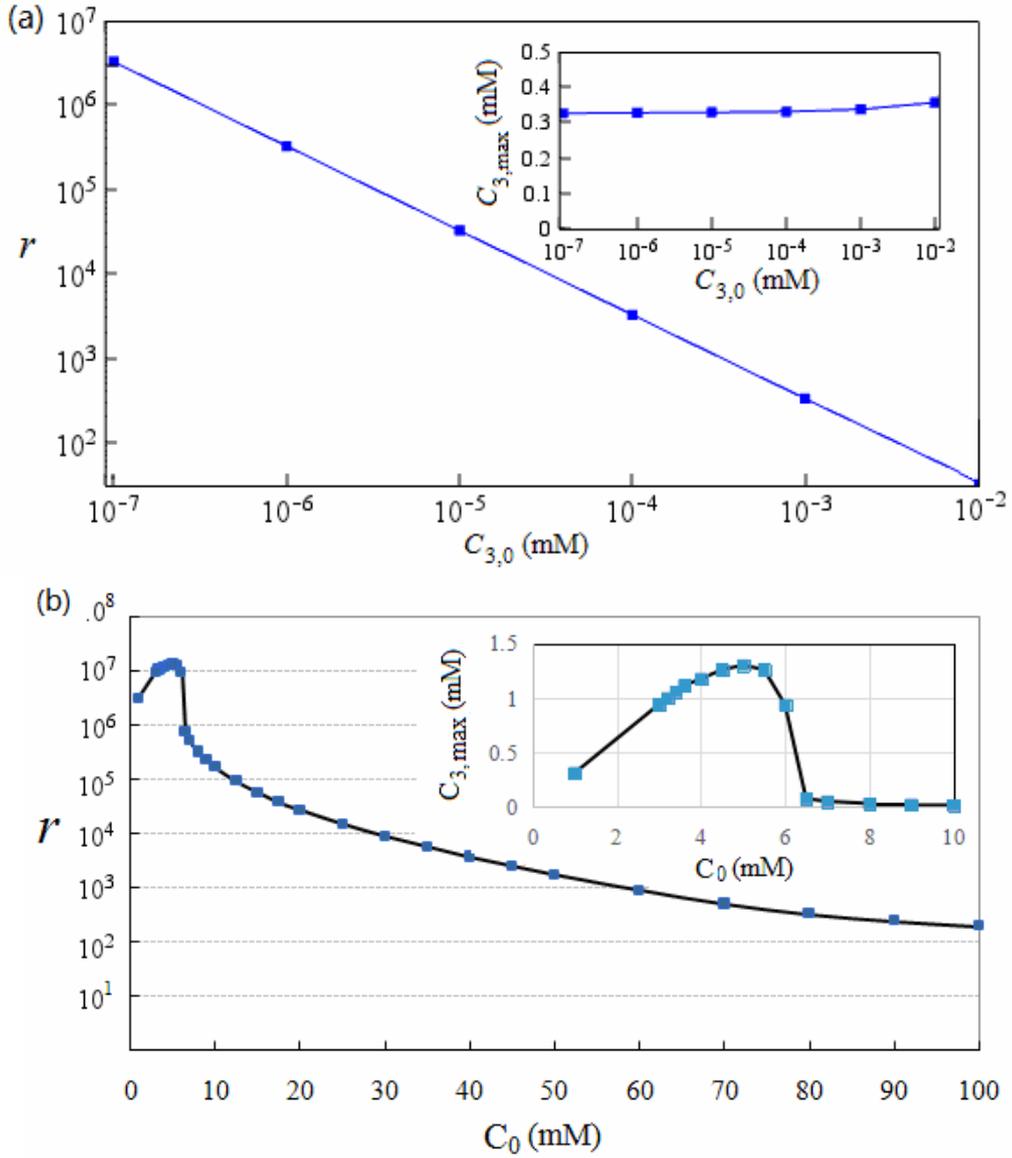

Fig. 5. Dependence of enrichment factor on (a) initial particle concentration $C_{3,0}$ and (b) the concentration of buffer ions ($C_0$), with insets showing the maximum particle concentration $C_{3,max}$ along the center line of microchannel.

Fig. 5b shows the dependence of limiting concentration on the buffer ion concentration. In this example, the initial concentration of particles is fixed at $C_{3,0} = 10^{-7}$ mM, while the buffer concentration $C_0$ changes from 1mM to 100mM under a constant $V_{cm} = 30 V_T$. It could be seen that for $C_0 \leq 5$mM, the particle enrichment factor grows with $C_0$, simply because that the upper limit defined by the concentration of $K^+$ is increased. However, beyond $C_0 = 5$mM, the enrichment factor decreases sharply. At $C_0 = 6.5$mM, the enrichment factor is less than 6% of its value at $C_0 = 0.5$mM (see the inset). After that, the enrichment factor decreases slowly. The reason behind these complicate behavior lies in the extent of ion depletion. For a given permselective membrane, the amount of counter ions that resides in pore spaces is almost constant (determined by net charges on surfaces of solid phase of membrane and neutrality condition, thus we use constant $C_1 = 2$mM with all $C_0$ values) and there is a limit for the amount of co-ions passing through the membrane under a fixed voltage. If the concentration of buffer ions is low, ion depletion is well developed and strong electric field is built as the energy barrier for charged particles. In this scenario, increasing buffer concentrations permits more particles to replace buffer co-ions, giving rise to higher enrichment factor. However, if the concentration of buffer ions is so high that ion depletion zone can not form sufficiently, the electric force that stops the flow of particles becomes weakened and the enrichment effect is compromised consequently. This result provides important insights into some of experimental observations. Specifically, in the early experiments of preconcentration using relatively pure solutions (e.g. fluorescent proteins and buffer ions), good preconcentration was observed. However, it is generally the case that preconcentration becomes less efficient when carried out over a complex buffer (e.g. serum), often with high-abundance background molecules (e.g. serum albumin or globulins). The result here suggests one mechanism to explain such behavior, since high abundance background ions and or molecules, will diminish the formation of the ion depletion and significantly reduce the ultimate possible concentration of interested particles in the device.

### D. Condition for particles to get enriched

To study behavior of particles of different sizes, we consider particles of the same charge ($Z_3 = -2$) but with diameter $\alpha$ times of Cl$^-$ ($D_3 = D_2/\alpha$). Fig. 6a shows the steady state concentration profiles of particles of different sizes under $V_{cm}=14V_T$ (the labels on the curves are values of $\alpha$). It is obvious that only particles with $\alpha > 2$ are enriched, with $\alpha = 3$ having the best enrichment effect. Particles of $\alpha \leq 2$ are not focused since their viscous drag is not strong enough to bring them into the microchannel with electroosmotic flow. Fig. 6b shows effects of particle charges. Here the size of particles is fixed ($D_3 = D_2/4$) and the valence $n$ (labeled on the curves) is changed. It is clear that only particles with charges $n < 4$ are focused, with $\alpha = 3$ having the highest enrichment factor.

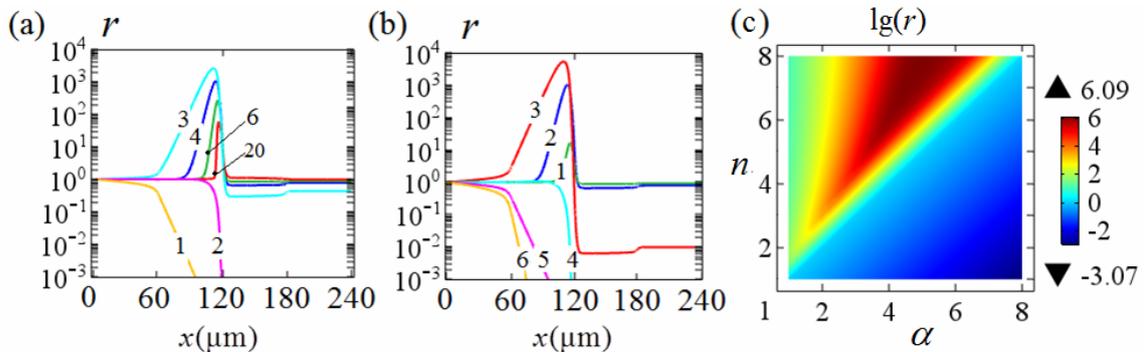

**FIG. 6.** Steady state concentration profiles of particles of (a) different sizes; (b) different charges; and (c) varied sizes and charges under $V_{cm}=14V_T$.

In principle, behaviors of particles in this system are governed by interplay between the leftward electrophoretic force ($\propto n$) and the rightward fluidic drag force ($\propto D_3^{-1}$ or $\alpha$). To permit a particle to enter the channel, magnitude of the viscous drag force must be greater than the electrophoretic force at the channel entrance. Because these two forces are almost equal for Cl$^-$ all over the channel (the flux of Cl$^-$ is close to zero, indicating a nearly zero migration speed and a balanced electrophoretic and drag force, c.f. Section H), the condition for the particle to enter the channel could be expressed as $n/\alpha < 1$. This condition can be justified by the results in Fig. 6a and 6b, where only particles with $n/\alpha < 1$ get enriched. More detailed study over this condition requires systematic changes of both $n$ and $\alpha$ in a continuous mode. For this purpose, Fig. 6c shows the maximum enrichment factor (colored in the order of 10) with varied values of $n$ and $\alpha$ (both from 1.0 to 8.0, at an interval of 0.2). It is clear that there is a partition line with $n/\alpha = 1$, with an enrichment factor of 1.0. Particles with $n/\alpha < 1$ are enriched ($r > 1$), while those with $n/\alpha < 1$ are not enriched ($r < 1$). Bearing in mind that $n/\alpha$ is proportional to the electrophoretic mobility of the particle and $n/\alpha = 1$ is representing buffer co-ion Cl$^-$, the condition for particle enrichment can be expressed as: *only particles with the electrophoretic mobility smaller than buffer co-ions can be focused*. It is noteworthy that although $n/\alpha < 1$ guarantees that particles can enter the microchannel with fluid flow, it does not mean that particles with smaller $n/\alpha$ will be *better* focused. In fact, particles with smaller charge or larger size are subject to weaker electric force or higher viscous drag forces, thus are easier to overcome the electric energy barrier before the ion depletion zone and leak into the downstream channel. Therefore, there is a maximum value of enrichment factor for situations of changing either particle size or electric charge, beyond which the enrichment factor of particles of smaller charges or larger particles are reduced (see Fig. 6).

### E. Replacement of buffer co-ions by focused particles

In contrast to particle concentration, which can be measured experimentally, concentrations of buffer ions in such micro-nano-fluidic channels have never been measured experimentally or studied numerically (in the presence of concentrating particles). In this end, Fig. 7a and 7b show the steady state concentration of K$^+$, Cl$^-$ and P$^{2-}$ under $V_{cm} = 14V_T$ and $V_{cm} = 34V_T$, respectively. In both cases, ion depletion occurs near the membrane location and concentrations of all the charged species are very low in downstream channel. Under low $V_{cm}$ (*e.g.* $14V_T$), concentration of P$^{2-}$ is significantly smaller than that of Cl$^-$ all over the channel (see Fig. 7a). In this scenario, the electric field and the electroosmotic flow are dominated by K$^+$ and Cl$^-$. However, if $V_{cm}$ is high (*e.g.* $34V_T$), K$^+$ ions are neutralized mainly by P$^{2-}$ in the focused region, leaving an almost-zero concentration for Cl$^-$, *i.e.* Cl$^-$ ions are almost completely replaced by P$^{2-}$ (see Fig. 7b). This therefore imposes an upper limit on the concentration of P$^{2-}$: it can not exceed that required to neutralize K$^+$. Another important finding from Fig. 7b is that the concentration of K$^+$ in the particle-focused region is non-negligibly lower than that in regions with low concentration of P$^{2-}$. For example, the concentration of K$^+$ under $V_{cm} = 34V_T$ is about 0.655mM in the replaced region, significantly lower than 1mM near the entrance. This variation is determined by the difference in mobilities of P$^{2-}$ and that of Cl$^-$, and the conservation of current. More specifically, near the entrance, concentration of P$^{2-}$ is small, current of negative charges is carried mainly by Cl$^-$ with high mobility. In the meantime near the membrane, this current is carried almost fully by P$^{2-}$ with low mobility. Conservation of the negative charge current requires that the electric field must be higher in the replaced region, so that the negative charges carried by different species have the same phenomenal migration speed. A higher electric field corresponds to a lower concentration of both $K^+$ and negatively charged species. Therefore, accompanying the replacement of anions by the focused particles, the concentration of buffer $K^+$ is reduced.

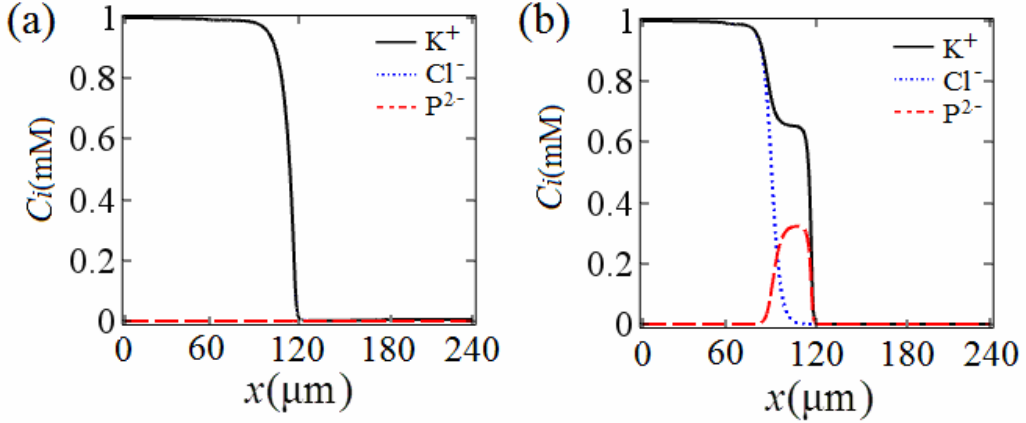

FIG. 7. Concentrations of $K^+$, $Cl^-$ and $P^{2-}$ under (a) $V_{cm}=14V_T$ and (b) $V_{cm}=34V_T$.

### F. Electric field

Enrichment of particles requires that fluid drag force be greater than the electrophoretic force at the entrance of microchannel, while in the mean time, the later must be greater than the former somewhere inside. This requirement has to be fulfilled by the variation of electric field[14], because the average fluid flow speed is constant all over the microchannel, determined by the conservation of fluid.

Fig. 8 gives profiles of electric potential $\Phi$ along the center line of the channel and the corresponding electric field component $E_x$ under varied $V_{cm}$. At low $V_{cm}$, values of $\Phi$ and $E_x$ are generally dominated by the channel geometry and values of $V_L$ and $V_R$. Increasing of $V_{cm}$ lowers $\Phi$ near the membrane and in the downstream microchannel, elevates $E_x$ in the upstream channel and decreases it in downstream. For example, at $V_{cm}=34V_T$, the maximum value of $E_x$ in the front of the depletion zone is ~542V/cm, nearly 25 times of the channel average value defined by $V_L$ and $V_R$ (21.5V/cm). In addition, under $V_{cm}>18V_T$ the direction of the electric field in the downstream microchannel is reversed ($E_x<0$, although its magnitude is relatively small. Further increase of $V_{cm}$ induces significantly stronger electric fields within the microchannel in the upstream, and outside the microchannel in the down stream. Reverse of electric field direction in downstream microchannel has been reported experimentally[14].

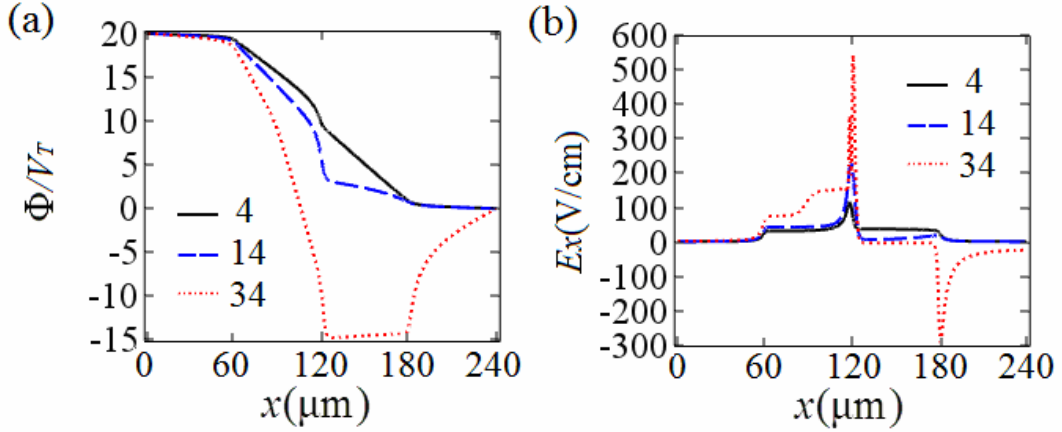

**FIG. 8.** The electric potential (a) and electric field (b) along the center line under varied $V_{cm}$.

It is noteworthy that the electric field in the particle focused region is much higher than that in the buffer ion dominated region under high cross membrane voltage $V_{cm}=34V_T$. This is a consequence of almost full replacement of buffer co-ions by the focused particles and simultaneous decreasing of counter ion concentration(c. f. Section E).

### G. Pumping effect

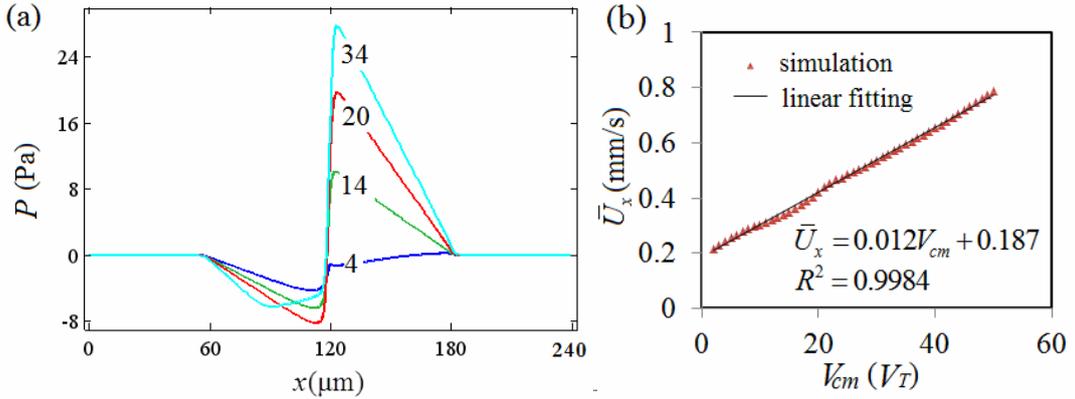

FIG. 9 (a) Pressure distribution along the center of the channel; (b) Dependence of the average flow speed on cross-membrane voltage

Accompanying ion depletion and formation of extended space charge near the membrane surface, the fluid nearby is accelerated and rotating vortices are generated under the amplified electric field. These vortices decrease the pressure in the upstream channel and increase the pressure in downstream, *i.e.* a fluid pumping effect is generated(see Fig. 9a). This effect has been studied qualitatively through experimental observations[14]. More insights into this effect could be obtained from the largely linear relationship between the average fluid flow speeds and $V_{cm}$ as shown in Fig. 9b. From the fitted expression, we may identify the contribution from the traditional EOF ($\bar{U}_x$ under $V_{cm}=0$) and that from the ICP effect (the difference between the actual $\bar{U}_x$ and that caused by EOF). For example, under $V_{cm}=50V_T$ the average fluid flow speed is 0.79mm/s. The contribution from ICP effect is about 3.4 times of that from EOF (0.18mm/s).

This pumping effect demonstrates an additional advantage of this system: it speeds up the particle preconcentration processes by several times.

## H. Fluxes of charged species

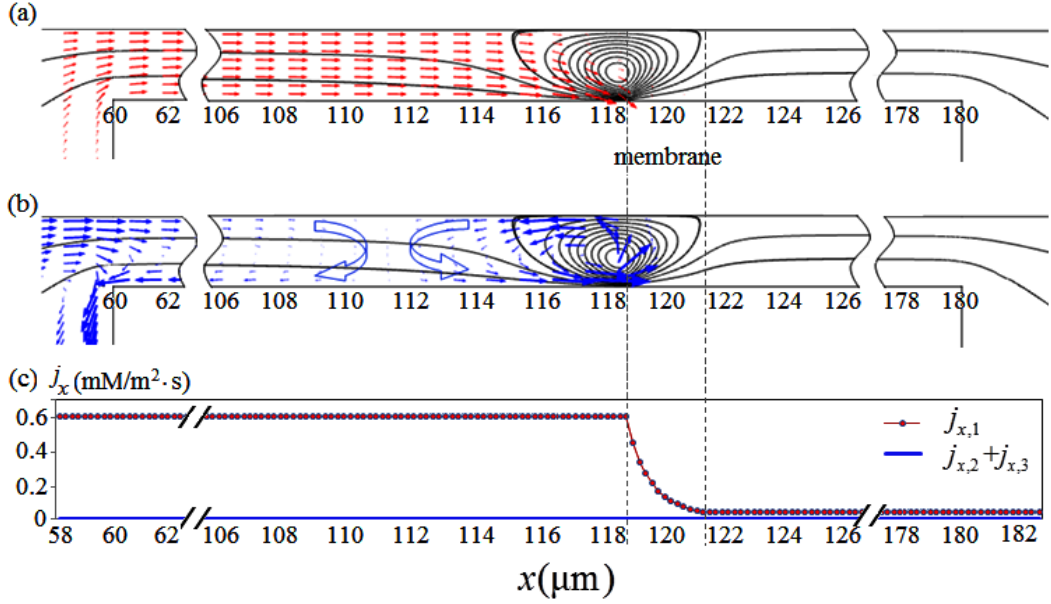

**FIG. 10.** (a) Flux field of $K^+$ under $V_{cm} = 10V_T$; (b) Total flux of $Cl^-$ and $P^{2-}$; (c) one dimensional flux description.

Fig. 10 depicts the steady state fluxes of $K^+$ (red arrows in Fig. 10a) and negatively charged species ($Cl^-$ and $P^{2-}$, blue arrows in Fig. 10b), along with the streamline of the fluid flow. For clarification purposes, fluxes for negatively charged species in Fig. 10b are magnified 8 times (as compared with those of $K^+$ in Fig.10a), because their magnitudes are small. In contrast to simple flux profile of $K^+$ that enter the microchannel and flow out through the membrane, flux field of negative species is complicated. There is a counter clockwise vortex influx of negative charged species near the membrane in the upstream channel (driven largely by fluidic vortex, shown in black color). While at the far end of upstream channel, there is a clockwise flux vortex as a result of balance between the nonuniform rightward fluidic flow and leftward electrophoretic motion. In the region where these two vortices meet, the fluxes of negative particles are almost zero. If we integrate the fluxes of each species over cross section (cs) of the channel $j_{x,i} = \int_{cs} J_{x,i} dy$, $i = 1, 2, 3$, we get a constant large flux for $K^+$ ($j_{x,1}$), and a constant near-zero flux for negatively charged species ($j_{x,2} + j_{x,3}$) in the upstream microchannel (see Fig. 10c). While, in the downstream, fluxes of all the charged species are almost zero. In other words, in one dimensional description, counter ions enter the microchannel and leave the channel through membrane, while negatively charged species are *almost* stationary at the steady state. These descriptions are also valid for time-dependant data, although replacement of $Cl^-$ by $P^{2-}$ takes place, until the steady state is reached.

## I. Mechanism picture of electric field driven preconcentration

Integration of results described above permits us to draw a full mechanism picture for transport of cations, anions, particles and the solvent fluid collectively in a micro-nano-fluidic channel with ICP. Application of $V_{cm}$ induces ion depletion that (1) desalinates the solution in the downstream; (2) stops the motion of co-ions and particles; and (3) speeds up the fluid flow as a pump. If the electrophoretic mobilities of the particles are smaller than the buffer co-ions, particles enter the microchannel with fluid and accumulate at the front of the depletion zone and replace co-ions. During this process, (1) fluid flows at a constant speed; (2) co-ions are almost stationary and get slowly replaced by focused particles; and (3) counter ions enter the microchannel and escape through the membrane at a high speed. In the mean time, replacing anions with charged particles decreases the concentration of cations, which further set limits for concentration of charged particles and anions. As a result, a plateau of particle concentration is formed and expands to the upstream channel under high $V_{cm}$. These particle accumulation and ion replacement processes proceed until a steady state is reached.

This mechanism picture may provide new, fundamental guidelines for researchers to explain their experimental observations. For example, one may think that behaviors of negatively charged particles and that of buffer anions should be similar in the sense of transport dynamics and/or enrichment. Simulation results here clarify these misunderstandings by highlighting diverse dynamics of various charged species (see Fig. 11) and totally different distributions of co-ions and the charged particles(see Fig. 7). These results tell us that one simply can not infer concentration of anions based on that of co-charged particles. The second common misunderstanding is that the buffer ions will get enriched, as speculated based on observation of focused particle concentration. The actual situation is that focusing of charged particles will reduce concentrations of both co-ions and counter ions (see Fig 7). Another common practice, to observe fluid vortices using charged fluorescent particles, is also not safe, because the shape of the vortical flow field of fluid is different from that of particle flux (see Fig. 10b).

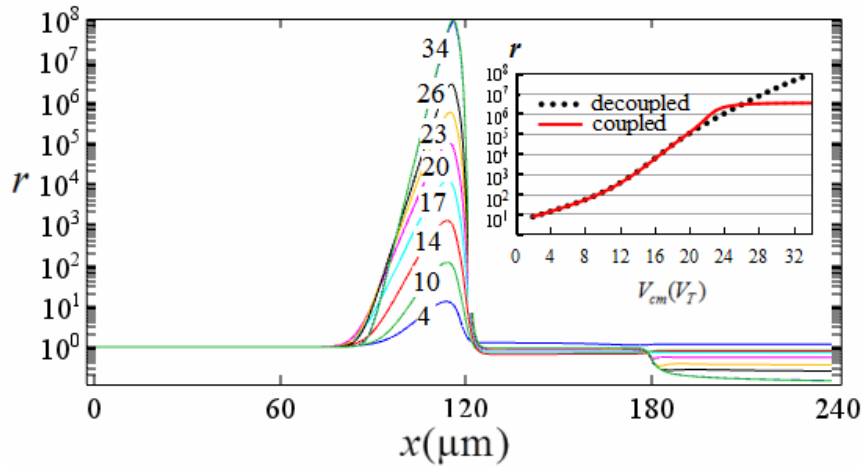

Fig. 11. Steady state profiles of particle concentration along the center line of the microchannel under varied $V_{cm}$ (labels on the curves are $V_{cm}$ in terms of $V_T$) using a decoupled model. Inset shows the compared the maximum enrichment factor from coupled and decoupled models.

In theoretical aspects, analyses of such systems have been often conducted through decoupling electrodynamics flow of electrolytes and convective diffusion of charged particles. This treatment may not work in most actual preconcentration situations because the basic assumption for such treatment, *i.e.* particle enrichment does not affect the fluid flow and distribution of buffer ions, becomes invalid when the charge carried by the focused particles is not negligible compared with that carried by buffer co-ions. For example, if we use decoupled model (to calculate fields of the fluid velocity, pressure and electric potential first without the presence of particle, and then use these fields to calculate the convective diffusion of charged particles in these constant fields) to calculate enrichment factor under varied cross membrane voltage (cases in Fig. 4), we will get an ever growing enrichment factor without upstream band expansion(see Fig. 11). As shown in the inset, the two systems produces almost identical results Under low $V_{cm} \leq 20 V_T$. However, the results of decoupled model under $V_{cm} > 20$ are completely unreasonable. For example, at $V_{cm} = 34 V_T$, the enrichment factor reaches $\sim 10^8$, corresponding to a particle concentration $\sim 10$ times higher than buffer ions. Because the enrichment factor at $V_{cm} = 20 V_T$ is $\sim 10^5$, corresponding to a particle concentration of ~0.01mM and charges carried by particles ~0.02 times of that of Cl$^-$. This offers us a simple *post priori* criterion for validity of the decoupled model: the charge carried by the focused particle should be less than ~2% of that carried by buffer co-ions. If a decoupled model yields a particle concentration with charges less than this value, the results are safe, otherwise they are not reliable and need revision. From these discussions, we know that charged particles in such systems are playing active roles in such systems through contribution to the charge density of the fluid and mediating the electric field. This may further affects the flow of fluids and transport of buffer ions. These fundamental understandings will significantly help future researches in analysis of experimental results or design of similar micro-fluidic systems.

## IV. CONCLUSION

This paper describes full mechanism picture of particle preconcentration micro-nano-fluidic system, through accurate multi-physics simulation. Dynamics of fluids and charged species, conditions of particle enrichment, replace of particles for buffer co-ions, and the pumping effects are elaborated. Condition for particles that may be focused is identified and validity of decoupled simulation model is discussed. Such fundamental knowledge provides new way of analyzing experimental results, especially when phenomena that are not directly observable (such as dynamics of buffer ions, and fluid-particle interactions) are involved. The results and findings of this simulation have great potential in guiding the device design and protocol development of all similar micro-nano-fluidic preconcentration systems, including single channel preconcentration [28-31], integration of particle preconcentration with subsequent separation *etc*[32-33].

ACKNOWLEDGMENT

This work is supported by the National Natural Science Foundation of China (Grant Nos. 11372229, 21576130, 21490584), and the State Key Laboratory of Materials-Oriented Chemical Engineering (KL 13-18). Z Li was also partially supported by SMART Center (BioSyM IRG) funded by National Research Foundation of Singapore, to enable this work..

# SUPPORTING INFORMATION

Full COMSOL (v5.2a) model is provided with all the necessary setup for steady state analyses. Users may run from study-1 (changing $V_{cm}$) to study-4 (changing c3,0) sequentially to obtain most results for steady state analyses in this paper.